\documentclass[conference]{IEEEtran}
\IEEEoverridecommandlockouts
\usepackage{amsmath,amssymb,amsfonts}
\usepackage{algorithmic}
\usepackage{graphicx}
\usepackage{textcomp}
\usepackage{xcolor}
\def\BibTeX{{\rm B\kern-.05em{\sc i\kern-.025em b}\kern-.08em
    T\kern-.1667em\lower.7ex\hbox{E}\kern-.125emX}}

\usepackage{amsmath,amssymb,amsfonts}
\usepackage{siunitx} 
\DeclareMathOperator{\Ima}{Im}
\DeclareMathOperator{\sign}{sign}

\usepackage{amsthm}

\usepackage{import} 

\usepackage{tikz}
\usepackage{pgfplots}
\pgfplotsset{compat=1.16}

\theoremstyle{definition}
\newtheorem{definition}{Definition}

\usepackage{caption} 
\usepackage{subcaption} 

\usepackage{tabularx}
\usepackage{booktabs}
\usepackage{color, colortbl}
\definecolor{Gray}{gray}{0.925}

\usepackage{xspace}
\makeatletter
\DeclareRobustCommand\onedot{\futurelet\@let@token\@onedot}
\def\@onedot{\ifx\@let@token.\else.\null\fi\xspace}
 
\def\ie{i.e\onedot} \def\Ie{\emph{I.e}\onedot}

\def\etal{\emph{et al}\onedot}
\makeatother

\usepackage{etoolbox}
\expandafter\patchcmd\csname citeauthor \endcsname
  {\begingroup}{\begingroup\aftergroup\@}{}{}

\usepackage{float}
\usepackage{pdfpages}
\usepackage{subcaption}

\usepackage{algorithmic}
\usepackage{graphicx}
\usepackage{textcomp}

\usepackage[hidelinks]{hyperref}
\usepackage{xcolor}
\hypersetup{
    colorlinks,
    linkcolor={red!50!black},
    citecolor={blue!50!black},
    urlcolor={blue!80!black}
}

\usepackage{cleveref}

\usepackage[maxnames=1, mincitenames=1,maxbibnames=100,sortcites=true,citestyle=numeric-comp,backend=biber,bibencoding=utf8,style=ieee]{biblatex}

\bibliography{dpsecure_private}

\usepackage[T1]{fontenc}
\newcommand\thefont{\expandafter\string\the\font}

\begin{document}


\title{Gradient Masking and the Underestimated Robustness Threats of Differential Privacy in Deep Learning}

\author{\IEEEauthorblockN{Franziska Boenisch, Philip Sperl, and Konstantin Böttinger}
\IEEEauthorblockA{
\textit{Fraunhofer Institute for Applied and Integrated Security}\\
\{franziska.boenisch; philip.sperl; konstantin.boettinger\}@aisec.fraunhofer.de}
}

\maketitle

\begin{abstract}
An important problem in deep learning is the privacy and security of neural networks (NNs). 
Both aspects have long been considered separately. 
To date, it is still poorly understood how privacy enhancing training affects the robustness of NNs. 
This paper experimentally evaluates the impact of training with Differential Privacy (DP), a standard method for privacy preservation, on model vulnerability against a broad range of adversarial attacks.
The results suggest that private models are less robust than their non-private counterparts, and that adversarial examples transfer better among DP models than between non-private and private ones. 
Furthermore, detailed analyses of DP and non-DP models suggest significant differences between their gradients.
Additionally, this work is the first to observe that an unfavorable choice of parameters in DP training can lead to gradient masking, and, thereby, results in a wrong sense of security. 
\end{abstract}


\section{Introduction}
\label{sec:introduction}
Deep neural networks (DNN) are applied in an increasing number of sensitive domains.
Yet, they exhibit vulnerabilities against a broad range of attacks threatening privacy and security \cite{Shokri.2017Membership,Fredrikson.2015Model,Szegedy.2014Intriguing}. 
To protect DNNs from such risks, in recent years, numerous defense strategies have been proposed.

As for privacy protection, \emph{Differential Privacy} (DP) \cite{Dwork.2006Differential} has become a standard framework.
By providing strong mathematical guarantees on privacy disclosure, DP allows to conduct data analyses on a population while protecting individual privacy.
With the introduction of the differential private stochastic gradient descent algorithm (DP-SGD) \cite{Abadi.10242016Deep}, DP has been made available for numerous machine learning (ML) algorithms, such as DNNs.
It has been shown that training with DP can mitigate the risks of privacy threatening attacks \cite{Shokri.2017Membership}.
At the same time, a large body of research has evolved around protecting DNNs against \emph{adversarial examples} \cite{Goodfellow.2014Explaining,Huang.2011Adversarial,Madry.2017Towards}.
Such data points contain small and human-imperceptible perturbations, forcing the attacked DNNs into misclassifications.
This poses a security threat for all neural network based architectures and depending on the use-case may lead to severe incidents.
Currently, adversarial retraining is considered the most effective method to protect against such attacks \cite{Ren.2020Adversarial}.
During this computationally expensive process, adversarial examples are used to retrain the DNNs using their original labels \cite{Madry.2017Towards}.

Even though the creation of models that are both private and secure at the same time poses a desirable goal, the majority of previous work focused on either one of the tasks \cite{Song.2019Membership}.
Only recently has the intersection of the different research branches received more attention \cite{Pinot.2019unified, Phan.2019Preserving, Phan.2019Heterogeneous, Phan.2019Scalable, Ibitoye.2021DiPSeN}.
The interrelation between privacy and security can be approached from two sides: either by evaluating the privacy implications of making a model more \emph{robust}, or by studying the influence of private training on model robustness.
So far, the former, namely the question of the influence of adversarial retraining on model privacy, has been investigated more thoroughly \cite{Giraldo.2020Adversarial,He.2020Robustness}.
It was shown that adversarial retraining decreases the membership privacy of a model's training data points. \Ie, 
for an attacker, it becomes easier to determine whether a specific data point was used during training \cite{Song.2019Membership,Mejia.2019Robust,Hayes.2020Provable}.
Regarding the latter perspective, first results suggest that applying DP-SGD to train a DNN has negative impacts on the model's robustness, even though there might potentially be privacy parameter combinations that are beneficial for both goals \cite{Tursynbek.2020Robustness}.
However, the experiments conducted to evaluate robustness in \cite{Tursynbek.2020Robustness} are limited to gradient-based adversarial attacks. 
Hence, it remains unclear whether these findings can be generalized and whether the robustness can be confirmed for other adversarial attack methods. 

To further shed light on the intersection between models' robustness and privacy, this work presents a comprehensive study building upon current findings.
In more detail, in addition to the previously evaluated robustness against gradient-based attacks \cite{Goodfellow.2014Explaining, Madry.2017Towards}, the experiments include gradient-free \cite{Brendel.2017Decision} and optimization-based methods \cite{Carlini.2017Towards}.  
Furthermore, to the best of the authors' knowledge, this work is the first one studying adversarial transferability \cite{Goodfellow.2014Explaining} among models with different privacy settings, and between private and non-private models.
Finally, it is also the first one analyzing DP models under the aspect of gradient masking \cite{Papernot.2017Practical} which might be the cause of a false sense of robustness.

In summary, the presented work makes the following contributions:
\begin{itemize}
    \item It experimentally evaluates the impact of DP-SGD training on adversarial robustness within a broad range of different privacy setups and parameters using various types of adversarial attacks. 
    Results suggest that DP models may be more prone to adversarial attacks than non-private models and that with increasing privacy through higher noise levels (especially in combination with high clip norms), decreasing adversarial robustness can be found.
    \item To exclude the effect of gradient masking from the robustness analyses, transferability attacks between private and non-private models are conducted.
    Their results suggest that adversarial examples transfer better from non-private models to private ones than the other way round.
    Additionally, there seems to be a higher transferability among DP models than among private and non-private ones.
    \item To investigate why DP models might be vulnerable to adversarial attacks, the gradients of the models are studied and compared with the gradients of non-private models.
    \item The work uncovers that in some settings, DP-SGD unintentionally causes gradient masking, which might explain the apparently increased robustness of certain settings reported in previous work~\cite{Tursynbek.2020Robustness}.
    Furthermore, it presents parameter combinations which intentionally provoke this effect.
    The findings are confirmed using a catalogue of criteria previously presented in literature \cite{Athalye.2018Obfuscated,Carlini.2019On}.
\end{itemize}




\section{Background and Related Work}
\label{sec:related}
This section depicts the theoretical background of deep learning classification, DP, adversarial machine learning (ML), and gradient masking.

\subsection{Deep Learning Classification}
\label{ssec:notation}
Let $f : \mathbb{R}^m \rightarrow \{1 \dots  k\}$ be a deep learning model for classification. 
The model $f$ maps the input $x$ to a discrete label $l \in \{1 \dots k \}$.
Internally, $f$ consists of $n$ parametric functions $f_i$ with $i \in \{1 \dots n \}$. 
Each function is represented by a layer of neurons that apply an activation function to the weighted output of the previous layer in order to create a new intermediate representation.
The layer $i$ is parameterized by a weight vector $\theta_i$, that incorporates the knowledge of $f$ learned during the training procedure \cite{Papernot.2017Practical}.

During training, $f$ is given a large number of input-label pairs (x,y) based on which the weight vectors $\theta_i$ are adapted.
Adaptation is usually performed based on backpropagation \cite{Chauvin.1995Backpropagation}.
Therefore, in a \emph{forward pass}, an input is propagated through the network by computing $f(x) = f_n (\theta_n, f_{n-1} (\theta_{n-1}, \dots f_2 (\theta_2, f_1 (\theta_1, x))))$. 
A loss function $\mathcal{L}$ then quantifies the difference between $f(x)$ and the original label $y$. 
To reduce the cost during training, in the \emph{backward pass}, the weight vectors $\theta_i$ are adjusted based on the \emph{gradients}, i.e. the first order partial derivatives, of each weight w.r.t the costs.

\subsection{Differential Privacy}
\label{ssec:rel_dp}
DP \cite{Dwork.2006Differential} is a mathematical framework that was introduced to provide privacy guarantees for algorithms analyzing databases.
Its definitions are described for neighboring data bases, i.e. data bases that differ in exactly one entry.
The $(\epsilon, \delta)$-variant \cite{Dwork.2013Algorithmic} is defined as follows.

\begin{definition}[$(\epsilon,\delta)$-Differential Privacy]
\label{def:edDP}
A randomized algorithm $\mathcal{K}$ with domain $\mathbb{N}^{|\mathcal{X}|}$ provides $(\epsilon,\delta)$-DP, if for all neighboring databases $D_1, D_2 \in \mathbb{N}^{|\mathcal{X}|}$ and all S $\subseteq\Ima(\mathcal{K})$
\begin{align}
\Pr[\mathcal{K}(D_1)\in S] \leq e^\epsilon \cdot \Pr[\mathcal{K}(D_2)\in S] + \delta \text{.}
\end{align}
\end{definition}

The parameters $\epsilon, \delta > 0$ can be interpreted as the privacy budget and the probability of violating this particular level of privacy, respectively.
The budget parameter $\epsilon$ sets an upper bound to possible privacy leaks. 
With a smaller privacy budget, higher levels of privacy are achieved.

\subsection{Differential Private Stochastic Gradient Descent}
\label{ssec:rel_dpsgd}
In deep learning, the randomized algorithm from the definition of DP usually refers to the network, while $D_1$ and $D_2$ are training data sets that differ in exactly one data point.
In order to integrate privacy as an optimization goal during the training process of deep learning models, DP-SGD \cite{Abadi.10242016Deep} was proposed.
The algorithm describes an adaptation of the original stochastic gradient descent algorithm \cite{Amari.1993Backpropagation}. 
Privacy is mainly achieved by two parameters, a noise value $\sigma$ and a gradient clipping bound $C$. 
After the gradients are computed with respect to the loss $\mathcal{L}$, their values are clipped to $C$ to reduce the influence of single samples. 
Then, Gaussian noise with zero mean and a variance of  $\sigma^2C^2\mathbb{I}$ is added to the clipped gradients in order to protect privacy.
For a detailed depiction of the algorithm, see \cite{Abadi.10242016Deep}.

\subsection{Adversarial Examples}
\label{ssec:rel_adv}
According to Szegedy \etal \cite{Szegedy.2014Intriguing}, adversarial examples can be characterized as follows:
Given an original sample $x$, the data point $x'$ is an adversarial example if (1) $x' = x + \beta$ for a small value $\beta$, and (2) $f(x) \neq f(x')$.
Hence, while $x'$ differs only slightly from $x$, the attacked model wrongly predicts the class of $x'$.
To measure the distance between the benign sample $x$ and its adversarial counterpart $x'$, usually an $l_p$-norm is used.
In literature, several methods have been proposed to generate such adversarial examples, which can be divided in gradient, optimization, and decision-based methods \cite{Zhang.2021Detecting}.
A summary of current attack methods can be found in the survey by Ren \etal~\cite{Ren.2020Adversarial}.
In the following, four widely used methods that were included in the experiments of this work are introduced.
Other important methods worth mentioning are DeepFool~\cite{MoosaviDezfooli.2016DeepFool} and the One-Pixel-Attack \cite{Su.2019One}.
The former method explores and approximates the decision boundary of the attacked model by iteratively perturbing the inputs. 
For the latter one, solely one pixel of the benign images is changed.

\subsubsection{Fast Gradient Sign Method}
\label{ssec:rel_fgsm}
The Fast Gradient Sign Method (FGSM) \cite{Goodfellow.2014Explaining} is a single-step, gradient-based method that relies on backpropagation to find adversarial examples. 
For the model parameters $\theta$, input $x$, target $y$, and loss function $\mathcal{L}(\theta,x,y)$, $x'$ is calculated as follows
\begin{align}
x' = x + \epsilon \sign(\nabla_x \mathcal{L}(\theta,x,y)) \text{.}
\end{align}

\subsubsection{Projected Gradient Descent}
\label{ssec:rel_pgd}
The Projected Gradient Descent (PGD) method \cite{Madry.2017Towards} represents a multi-step variant of FGSM.
Given a distance norm $p$, $x'_0$ is initialized in an $L_p$ ball around the original sample, iteratively adapted as in FGSM, and if necessary, the perturbation is projected back into the $L_p$ ball.
Starting with $x'_0$, a step size $\alpha$, and projection $\Pi_{L_p}$, the adversarial example $x'_t$ at iteration $t$ is constructed as follows
\begin{align}
x'_t = \Pi_{L_p}(x'_{t-1} + \epsilon \sign(\nabla_x \mathcal{L}(\theta,x'_{t-1},y)) \text{.}
\end{align}

\subsubsection{Carlini and Wagner}
\label{ssec:rel_cuw}
The optimization-based Carlini and Wagner (CW) method \cite{Carlini.2017Towards} uses a loss function that optimizes the distance between a target class $y'$ and the most likely class.
Given a model $f$ with logits $Z$, the CW$_2$ attack can be formulated as follows:
\begin{align}
\min \parallel x' - x \parallel_2^2 + c \cdot \mathcal{L}(\theta,x',y)
\end{align}
with the loss function being
\begin{align}
\mathcal{L}(\theta,x',y) = \max (\max\{Z(x')_i :  i \neq t \} -Z(x')_t -k )   \text{.}
\end{align}
The constant $c$ is determined by binary search. 

\subsubsection{Boundary Attack}
\label{ssec:rel_boundary}
The Boundary Attack (BA$_2$) \cite{Brendel.2017Decision} is a multi-step method that relies solely on the model's decision to generate adversarial examples.
Therefore, $x'$ is initialized as an adversarial point, \ie, $f(x)\neq f(x')$. 
Afterwards, a random walk on the boundary between the adversarial and the non-adversarial region is performed in order to minimize the distance between $x$ and $x'$, while keeping $x'$ adversarial. 
With the BA$_2$ method, adversaries are able to craft adversarial examples even if the gradients of the attacked NN are not available.

\subsection{Adversarial Transferability}
\label{ssec:rel_transfer}
It has been shown that adversarial examples transfer between models \cite{Goodfellow.2014Explaining}, \ie adversarial examples that are generated on one model are often successful in fooling other similar models.
The degree of success depends on several factors, among which the complexity of the model that the adversarial examples are crafted on, and the similarity between both models \cite{Demontis.2019Why}.

\subsection{Privacy and Robustness}
\label{ssec:priv_rob}
Research suggests that the goals of achieving privacy and robustness in ML models are not always in line.
It was found that adversarial retraining, \ie retraining a model with adversarial examples, increases membership privacy risks \cite{Song.2019Membership, Mejia.2019Robust}. 
With decreased membership privacy, it might be easier for an attacker to determine whether a specific data point has been used for model training \cite{Shokri.2017Membership}.
This negative impact of adversarial retraining on models' privacy can be explained with overfitting. 
The retraining might enforce the training data points more strongly to the model \cite{Hayes.2020Provable} leading to a decrease of the membership privacy \cite{Yeom.2018Privacy}.

Examined from the opposite perspective, it was shown that the noise of DP training can be exploited to craft adversarial examples more successfully \cite{Giraldo.2020Adversarial}.
In the work that is closest to the present, Tursynbek \etal \cite{Tursynbek.2020Robustness} observed that DP training can have a negative influence on adversarial robustness. 
Yet, the authors identified some DP parameters that give the impression of improved robustness against gradient-based adversarial example crafting methods.

Some research has already been dedicated to aligning model privacy and robustness.
Pinot \etal \cite{Pinot.2019unified} showed that model robustness and DP share similarities in their goals.
Furthermore, several mechanisms to integrate (provable) adversarial robustness into DP training have been proposed \cite{Phan.2019Heterogeneous, Phan.2019Preserving, Phan.2019Scalable, Ibitoye.2021DiPSeN}.

\subsection{Gradient Masking}
\label{ssec:rel_grad}
The term of \emph{gradient masking} \cite{Papernot.2017Practical} refers to adversarial defence methods intentionally or unintentionally reducing the usefulness of the protected model's gradients.
Thereby, gradient-based adversarial attacks are less successful.
As the models are often still vulnerable to non-gradient-based attacks, gradient masking results in a false sense of robustness.
The term \emph{gradient obfuscation} \cite{Athalye.2018Obfuscated} describes a specific case of gradient masking, in which defenses against adversarial attacks are designed such that they cause gradient masking.
Hence, the resulting level of robustness provided by such methods is often overestimated.
Athalye \etal \cite{Athalye.2018Obfuscated} identified three types of gradients with reduced utility:
\begin{enumerate}
    \item \emph{Shattered gradients} are non-existent or incorrect gradients caused either intentionally through the use of operations that are non-differentiable or unintentionally through numerical instabilities.
    \item \emph{Stochastic gradients} are gradients the depend on test-time randomness.
    \item \emph{Vanishing or exploding gradients} are gradients that are too small or too large to be useful for the computations performed during attacks.
\end{enumerate}

For models, potentially in combination with some defense method, the following properties indicate sane and un-masked gradients \cite{Athalye.2018Obfuscated, Carlini.2019On}:
\begin{enumerate}
    \item Iterative attacks perform better than one-step attacks.
    \item White-box attacks preform better than black-box attacks.
    \item Gradient-based attacks perform better then gradient-free attacks.
    \item Unbounded attacks should reach a 100\% adversarial success rate.
    \item Increasing the iterations within an attack should increase the adversarial success rate.
    \item Increasing the distortion bound on the adversarial examples should increase the adversarial success rate.
    \item White-box attacks perform better than transferability attacks using a similar substitute model.
\end{enumerate}
Violation of these criteria can be an indicator for masked gradients either within the model itself or due to the introduced defense method.


\section{Method and Experimental Setup}
\label{sec:setup}
In this paper, the intersection of privacy and security in DNNs is investigated.
More specifically, the impact of using a state-of-the-art training method to increase ML models' privacy levels, namely the DP-SGD optimizer, on model robustness is examined. 
Previous work suggests that using this optimizer with certain parameters might have positive effects on the robustness of the trained DNNs \cite{Tursynbek.2020Robustness}.
The experiments of this work show that this first indication of increased robustness may be due to masked gradients.
To validate this hypothesis, extensive experiments using two model architectures and four attack methods were conducted.
As can be seen in previous guidelines on the evaluation of adversarial robustness \cite{Athalye.2018Obfuscated, Carlini.2019On}, the selection of used attack methods heavily influences the perception of robustness.
Therefore, in the following, the experimental setup used to shed more light on the robustness of privately trained DNNs is summarized.

The experiments in this paper were conducted using the MNIST dataset \cite{LeCun.2010MNIST} consisting of 60000 train and 10000 test gray-scale images of size 28 × 28 pixel.

Two different model architectures were evaluated: an adaptation of the \emph{LeNet} architecture \cite{LeCun.1989Backpropagation} (1), and a \emph{custom} conv-net architecture (2), both depicted in Table \ref{tab:architectures}.
The networks were implemented using TensorFlow \cite{Abadi.2016TensorFlow} version 2.4.1 and trained for 50 epochs using a learning rate of 0.05, and a batch size of 250.
Stochastic gradient descent and TensorFlow Privacy's \cite{Googleetal..2018TensorFlow} implementation of the DP-SGD were used as optimizers to train the models without and with privacy, respectively.
The privacy parameters $\sigma$ and $C$ of the DP-SGD were specified during the experiments. 
In accordance to \cite{Tursynbek.2020Robustness} the value ranges were $\sigma \in \{0,1.3,2,3\}$ and $C\in \{1,3,5,10\}$.

The adversarial examples were generated using the Foolbox framework \cite{Rauber.2017Foolbox}.
As shown in \Cref{ssec:rel_adv} PGD$_{\infty}$, CW$_{2}$, and BA$_{2}$ were used as examples of a multi-step gradient-based, multi-step optimization-based, and a multi-step gradient-free attack, respectively. 
For every experiment, 1000 adversarial examples were generated based on 1000 random test data points that were predicted correctly by the model under attack.

\begin{table*}[t]
    \centering
    \begin{tabular}{cc}
    \toprule
    LeNet architecture & Custom architecture \\
    \midrule 
        Conv(f=6, k=(3,3), s=1, p=valid, act=relu)   &  Conv(f=16, k=(8,8), s=2, p=same, act=relu)\\
        2D Average Pooling(pool size=(2,2), s=1, p=valid) & 2D Max Pooling(pool size=(2,2), s=1, p=valid)\\
        Conv(f=16, k=(3,3), s=1, p=valid, act=relu) & Conv(f=32, k=(4,4), s=2, p=valid, act=relu)\\
        2D Average Pooling(pool size=(2,2), s=1, p=valid) & 2D Max Pooling(pool size=(2,2), s=1, p=valid)\\
        Flatten & Flatten\\
        Dense(n=120, act=relu) & Dense(n=32, act=relu)\\
        Dense(n=84, act=relu) & Dense(n=10, act=None)\\
        Dense(n=10, act=None) & \\
    \bottomrule
    \end{tabular}
    \caption{Architectures of the models used in the experiments. f: number of filters, k: kernel size, s: stride, p: padding act: activation function, n: number of neurons.}
    \label{tab:architectures}
\end{table*}

\section{Experimental Robustness Evaluation of DP models}
\label{sec:robustness}
This section describes the results of the robustness evaluation on different DP models for the respective attacks.

\subsection{Robustness Evaluation with the PGD$_{\infty}$ Attack}
\label{ssec:results_linfpgd}

\begin{figure*}
    \centering
    \begin{subfigure}[b]{\textwidth}
         \centering
         \input{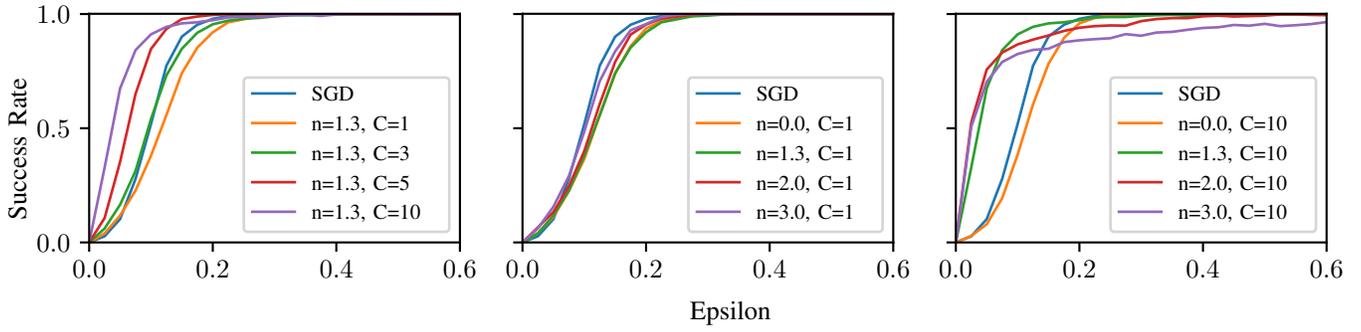}
         \caption{Custom architecture.}
         \label{fig:exp01_custom}
     \end{subfigure}
        \vfill
     \begin{subfigure}[b]{\textwidth}
         \centering
         \input{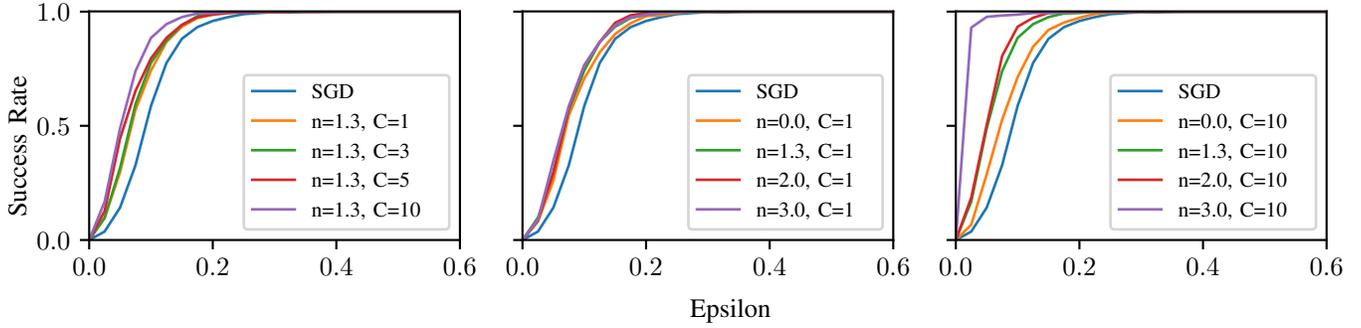}
         \caption{LeNet architecture.}
         \label{fig:exp01_lenet}
     \end{subfigure}
    \caption{PGD$_{\infty}$: adversarial success rate plotted against adversarial perturbation $\epsilon$.}
    \label{fig:exp01}
\end{figure*}

\paragraph{Perturbation-based Analysis}
In the first experiment, both the custom and the LeNet model were attacked using PGD$_{\infty}$. 
As this method creates bounded adversarial examples, the robustness was quantified using the adversarial success rate depending on the maximal adversarial perturbation budget $\epsilon$.
The perturbation budget was increased successively from $0.0$ to $0.5$ in steps of size $0.025$ for a fixed number of 40 attack iterations.
At every perturbation value, the adversarial success rate was measured.
Higher success rates for the same perturbation budget suggest a lower model robustness.
See Figure \ref{fig:exp01} for the results. 
For both model architectures and all privacy parameter combinations, an increase of the perturbation budget $\epsilon$, resulted in an increase of the success rate. 

For the custom architecture, the DP models with noise $\sigma=1.3$ and clip norm $C=1$ or $C=3$ achieve higher or similar levels of robustness compared to the non-private baseline model.
In contrast, models with higher clip norms can be attacked more successfully (see Figure \ref{fig:exp01_custom}).
The observation that some DP parameter combinations might be beneficial for robustness is in line with the findings by Tursynbek \etal~\cite{Tursynbek.2020Robustness} on their custom architecture.
Yet, this behavior cannot be observed for the LeNet model (see Figure \ref{fig:exp01_lenet}).
In this experiment, the adversarial success rate when attacking the non-private baseline model is lower than the success rate observed for every DP parameter combination. 

Similar to Tursynbek \etal~\cite{Tursynbek.2020Robustness}, a plateau-like stagnation of the success rate for combinations with higher noise of $\sigma=2$ or $\sigma=3$ in combination with a high clip norm of $C=10$ can be observed for the custom model.
This might be interpreted as a robustness improvement using these settings. 
However, in the LeNet model, no similar behavior can be observed.
Figure \ref{fig:exp01_lenet} (right) instead suggests that for $C=10$, the higher the amount of noise, the lower the model robustness.

For both model architectures, the success rates of different noise values in combination with a small clip norm of $C=1$ are very similar.
While for the custom architecture, the attacks achieve a higher success rate for the non-DP model, for the LeNet models, the opposite applies.

\begin{figure*}
    \centering
    \begin{subfigure}[b]{\textwidth}
         \centering
         \input{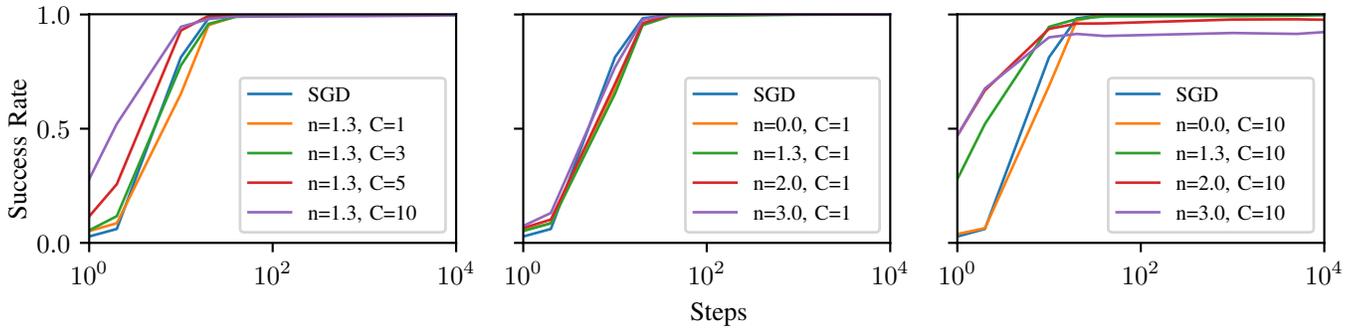}
         \caption{Custom architecture.}
         \label{fig:exp03_custom}
     \end{subfigure}
        \vfill
     \begin{subfigure}[b]{\textwidth}
         \centering
         \input{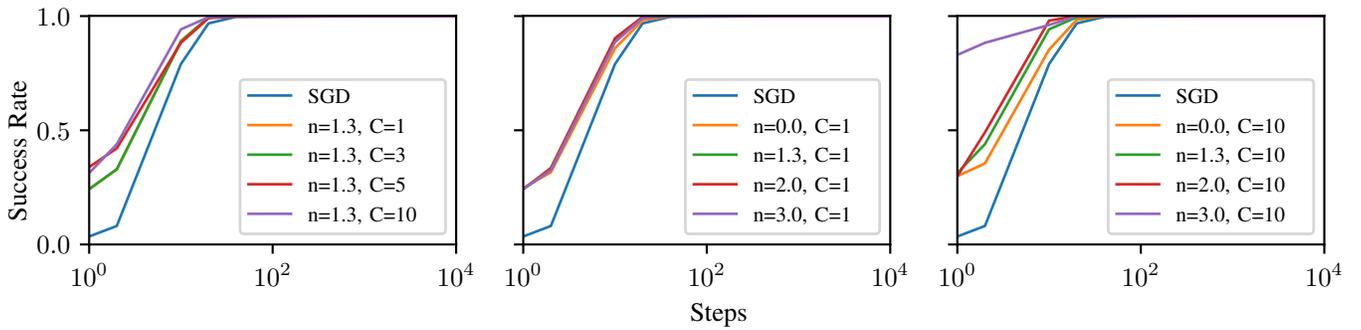}
         \caption{LeNet architecture.}
         \label{fig:exp03_lenet}
     \end{subfigure}
    \caption{PGD$_{\infty}$: adversarial success rate plotted against number of iterations.}
    \label{fig:exp03}
\end{figure*}

\paragraph{Step-based Analysis}
In the second experiment, the success rate of PGD$_{\infty}$ when attacking both architectures was evaluated depending on the number of iterations or so-called \emph{steps} for a fixed $\epsilon$ of \num{0.3}. 
The results suggest that, in general, the success rate increases with increasing numbers of iterations and finally reaches 100\% (see Figure \ref{fig:exp03}). 
However, in the custom model, with $\sigma=2, \sigma=3$, and $C=10$, the success rate reaches a plateau after \raisebox{-0.5ex}{\~{}}10 iterations and does not reach 100\%.
In contrast, for the LeNet architecture, no plateaus are reached and for the non-DP model highest robustness is reached with a sufficient amount of performed attack steps.

In summary, the experiments presented above confirm the findings by Tursynbek \etal \cite{Tursynbek.2020Robustness}.
Adversaries using PGD$_{\infty}$ cannot successfully attack their custom and privately trained model with a success rate of 100\%.
Based on this finding Tursynbek~\etal concluded, that using the DP-SGD optimizer during training simultaneously improves privacy, as well as robustness.
Yet, the experiments presented in this section using the LeNet model already suggest, that this finding does not generalize to other architectures.
Furthermore, in the next section, additional evidence will be presented showing that the evaluated private models do not show an increased level of robustness compared to their normally trained counterparts.

\subsection{Robustness Evaluation with the BA$_2$ Attack}
\label{ssec:results_boundary}
To further investigate the impact of DP-SGD training on model robustness, the gradient-free BA$_{2}$ was used to attack the DP models. 
Table \ref{tab:success_boundary} depicts the results for adversarial perturbations of $\epsilon=1$, and $\epsilon=2$ against the custom and LeNet architectures.
The attack was executed with \num{25000} iterations. 
For more iterations, no increase in success rates could be observed.

For both model architectures, the results suggest that increasing the clip value or the amount of noise for a high clip value leads to increased adversarial vulnerability.
The custom models with $\sigma=2$ or $\sigma=3$ and $C=10$, that reached a plateau in the adversarial success rate for PGD$_{\infty}$, are most vulnerable to BA$_{2}$ among all settings and for both model architectures. 
Solely for the parameter combinations of $C=1$ or $\sigma=0$ the BA$_{2}$ attack with perturbation $\epsilon=2$ reaches a lower success rate than for the non-DP baselines.
However, in general, the success rates on the DP models are higher than on the non-DP ones.

In accordance to the findings of the previous section, the experiments here again show, that DP models are generally not more robust than their normally trained counterparts.
The experiments even show that for certain parameter combinations the DP models are attacked even more easily using the BA$_{2}$ method.

\subsection{Robustness Evaluation with CW$_{2}$ Attack}
\label{ssec:results_cw}

In the final experiment, the optimization-based CW$_{2}$ attack was used.
This method generates adversarial examples in an unbounded manner.
Hence, to determine the robustness towards CW$_{2}$, the adversarial perturbation needed to achieve a 100\% adversarial success rate is measured.
Table~\ref{tab:cw_epsilon_rates} depicts the results of the experiments after 10,000 attack iterations.

The values suggest that with increasing noise, or with increasing clip norm, the amount of perturbation needed to achieve a 100\% success rate decreases. 
This indicates decreased adversarial robustness. 
For the LeNet architecture, the CW$_{2}$ attack requires less perturbations on all DP models than on the non-private baseline models.
The required perturbation budget decreases monotonically when increasing the clip value or noise.
Interestingly, in the custom architecture with $C=1$, or $\sigma=0, C=10$, a higher perturbation than in the non-private setting is required. 
Also, when setting the clip value to $C=10$, and varying the amount of noise, $\sigma=3$ requires higher perturbation than $\sigma=2$, or $\sigma=1.3$, hence, no monotonic decrease could be observed.

This experiment again underlines, that DP models do not generally exhibit a higher level of robustness.
First, the CW$_{2}$ attack is capable of generating adversarial examples with a success rate of 100\%.
Second, the required perturbation budget to generate the adversarial examples is in the majority of cases smaller for the DP models, than for the normally trained ones.

\begin{table}[t]
    \centering
    \begin{tabular}{lcc}
    \toprule
    Parameters & LeNet & Custom \\
    \midrule 
    SGD & $\epsilon=1.21$ & $\epsilon=1.20$\\
    \midrule 
    $\sigma=1.3, C=1$ &	$\epsilon=1.06$ & $\epsilon=1.37$\\ 
    $\sigma=1.3, C=3$ & $\epsilon=1.03$ & $\epsilon=1.17$\\ 
    $\sigma=1.3, C=5$ & $\epsilon=0.93$ & $\epsilon=0.80$\\ 
    $\sigma=1.3, C=10$ & $\epsilon=0.83$ & $\epsilon=0.45$ 	\\ 
    \midrule 
    $\sigma=0, C=1$  & $\epsilon=1.11$& $\epsilon=1.41$	\\ 
    $\sigma=1.3, C=1$  & $\epsilon=1.06$& $\epsilon=1.37$\\ 
    $\sigma=2, C=1$ & $\epsilon=1.04$& $\epsilon=1.31$ \\ 
    $\sigma=3, C=1$  & $\epsilon=1.01$ & $\epsilon=1.24$\\ 
    \midrule 
    $\sigma=0, C=10$ & $\epsilon=1.12$ & $\epsilon=1.38$	\\ 
    $\sigma=1.3, C=10$ & $\epsilon=0.83$ & $\epsilon=0.45$\\ 
    $\sigma=2, C=10$ & $\epsilon=0.63$ & $\epsilon=0.29$\\ 
    $\sigma=3, C=10$ & $\epsilon=0.15$ & $\epsilon=0.50$ 	\\ 
    \bottomrule
    \end{tabular}
    \caption{Adversarial perturbation $\epsilon$ required to achieve a 100\% adversarial success rate within \num{10000} iterations of Carlini and Wagner attack, rounded to two decimal points.}
    \label{tab:cw_epsilon_rates}
\end{table}

\begin{table}[t]
    \centering
    \begin{tabular}{lcccc}
    \toprule
    Parameters & \multicolumn{2}{c}{LeNet}  & \multicolumn{2}{c}{Custom}\\
     & $\epsilon=1$ & $\epsilon=2$  & $\epsilon=1$&$\epsilon=2$\\
    \midrule 
    SGD &  27.5\%& 79.4\%&	 19.1\%& 83.2\%\\
    \midrule 
    $\sigma=1.3, C=1$ &	 38.5\%& 75.7\%&	 21.6\%& 70.9\%\\ 
    $\sigma=1.3, C=3$ & 41.3\%& 81.3\%&	 26.6\%& 77.6\%\\ 
    $\sigma=1.3, C=5$ &  49.7\%& 82.1\%&	 51.3\%& 94.2\%\\ 
    $\sigma=1.3, C=10$ &  57.2\%&  	89.3\%&	 86.9\%& 99.9\%\\ 
    \midrule 
    $\sigma=0, C=1$ &  38.0\%& 	71.3\%&	 19.5\%& 65.2\%\\ 
    $\sigma=1.3, C=1$ &  38.5\%& 75.7\%&	 21.6\%& 70.9\%\\ 
    $\sigma=2, C=1$ &  36.3\%& 76.7\%&	 22.6\%& 72.4\%\\ 
    $\sigma=3, C=1$ &  40.9\%&  76.3\%&	 26.0\%& 74.1\%\\ 
    \midrule 
    $\sigma=0, C=10$ &  41.2\%& 	77.9\%&	 17.0\%& 71.7\%\\ 
    $\sigma=1.3, C=10$ &  57.2\%& 89.3\%&	 86.9\%& 99.9\%\\ 
    $\sigma=2, C=10$ & 65.2\%& 94.7\%&	 98.3\%& 100.0\%\\ 
    $\sigma=3, C=10$ &  91.5\%&  	91.8\%&	 93.6\%& 100.0\%\\ 
    \bottomrule
    \end{tabular}
    \caption{Success Rates of BA$_2$ with different perturbation values on both model architectures.}
    \label{tab:success_boundary}
\end{table}

\section{DP and Transferability}
\label{sec:transferability}
\begin{figure}
\begingroup%
\makeatletter%
\begin{pgfpicture}%
\pgfpathrectangle{\pgfpointorigin}{\pgfqpoint{2.690037in}{2.502909in}}%
\pgfusepath{use as bounding box, clip}%
\begin{pgfscope}%
\pgfsetbuttcap%
\pgfsetmiterjoin%
\definecolor{currentfill}{rgb}{1.000000,1.000000,1.000000}%
\pgfsetfillcolor{currentfill}%
\pgfsetlinewidth{0.000000pt}%
\definecolor{currentstroke}{rgb}{1.000000,1.000000,1.000000}%
\pgfsetstrokecolor{currentstroke}%
\pgfsetdash{}{0pt}%
\pgfpathmoveto{\pgfqpoint{0.000000in}{0.000000in}}%
\pgfpathlineto{\pgfqpoint{2.690037in}{0.000000in}}%
\pgfpathlineto{\pgfqpoint{2.690037in}{2.502909in}}%
\pgfpathlineto{\pgfqpoint{0.000000in}{2.502909in}}%
\pgfpathclose%
\pgfusepath{fill}%
\end{pgfscope}%
\begin{pgfscope}%
\definecolor{textcolor}{rgb}{0.000000,0.000000,0.000000}%
\pgfsetstrokecolor{textcolor}%
\pgfsetfillcolor{textcolor}%
\pgftext[x=0.314658in,y=1.951970in,left,base]{\color{textcolor}\rmfamily\fontsize{9.000000}{10.800000}\selectfont CW}%
\end{pgfscope}%
\begin{pgfscope}%
\pgfpathrectangle{\pgfqpoint{0.615180in}{1.803453in}}{\pgfqpoint{0.429317in}{0.429317in}}%
\pgfusepath{clip}%
\pgfsys@transformshift{0.615180in}{1.803453in}%
\pgftext[left,bottom]{\pgfimage[interpolate=true,width=0.430000in,height=0.430000in]{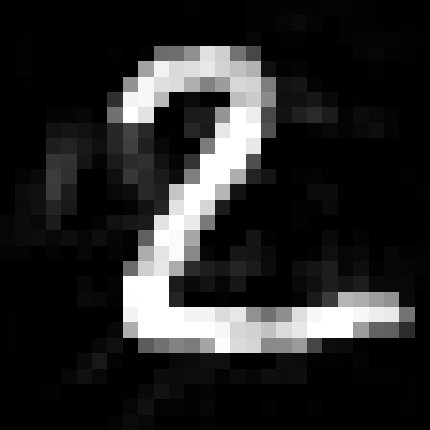}}%
\end{pgfscope}%
\begin{pgfscope}%
\definecolor{textcolor}{rgb}{0.000000,0.000000,0.000000}%
\pgfsetstrokecolor{textcolor}%
\pgfsetfillcolor{textcolor}%
\pgftext[x=0.829839in,y=2.316103in,,base]{\color{textcolor}\rmfamily\fontsize{9.000000}{10.800000}\selectfont pred: 0}%
\end{pgfscope}%
\begin{pgfscope}%
\pgfpathrectangle{\pgfqpoint{1.130360in}{1.803453in}}{\pgfqpoint{0.429317in}{0.429317in}}%
\pgfusepath{clip}%
\pgfsys@transformshift{1.130360in}{1.803453in}%
\pgftext[left,bottom]{\pgfimage[interpolate=true,width=0.430000in,height=0.430000in]{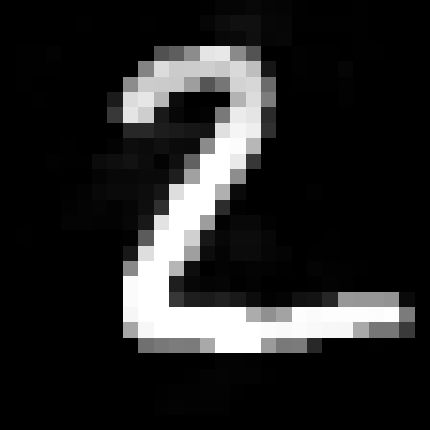}}%
\end{pgfscope}%
\begin{pgfscope}%
\definecolor{textcolor}{rgb}{0.000000,0.000000,0.000000}%
\pgfsetstrokecolor{textcolor}%
\pgfsetfillcolor{textcolor}%
\pgftext[x=1.345019in,y=2.316103in,,base]{\color{textcolor}\rmfamily\fontsize{9.000000}{10.800000}\selectfont pred: 6}%
\end{pgfscope}%
\begin{pgfscope}%
\pgfpathrectangle{\pgfqpoint{1.645540in}{1.803453in}}{\pgfqpoint{0.429317in}{0.429317in}}%
\pgfusepath{clip}%
\pgfsys@transformshift{1.645540in}{1.803453in}%
\pgftext[left,bottom]{\pgfimage[interpolate=true,width=0.430000in,height=0.430000in]{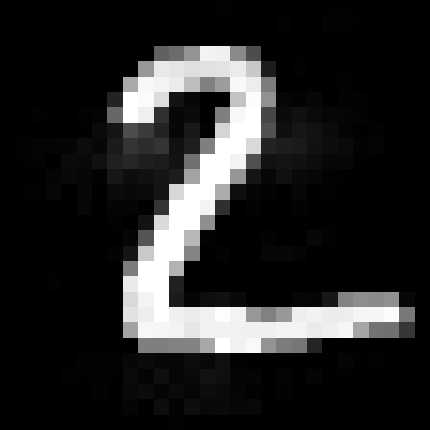}}%
\end{pgfscope}%
\begin{pgfscope}%
\definecolor{textcolor}{rgb}{0.000000,0.000000,0.000000}%
\pgfsetstrokecolor{textcolor}%
\pgfsetfillcolor{textcolor}%
\pgftext[x=1.860199in,y=2.316103in,,base]{\color{textcolor}\rmfamily\fontsize{9.000000}{10.800000}\selectfont pred: 8}%
\end{pgfscope}%
\begin{pgfscope}%
\pgfpathrectangle{\pgfqpoint{2.160721in}{1.803453in}}{\pgfqpoint{0.429317in}{0.429317in}}%
\pgfusepath{clip}%
\pgfsys@transformshift{2.160721in}{1.803453in}%
\pgftext[left,bottom]{\pgfimage[interpolate=true,width=0.430000in,height=0.430000in]{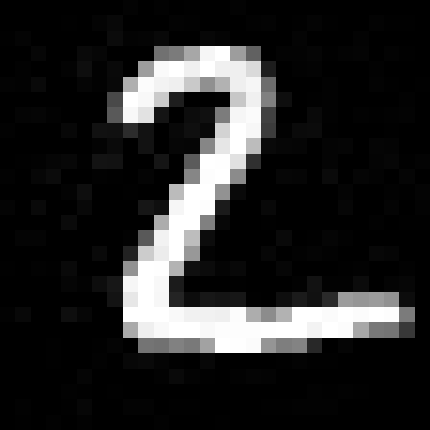}}%
\end{pgfscope}%
\begin{pgfscope}%
\definecolor{textcolor}{rgb}{0.000000,0.000000,0.000000}%
\pgfsetstrokecolor{textcolor}%
\pgfsetfillcolor{textcolor}%
\pgftext[x=2.375379in,y=2.316103in,,base]{\color{textcolor}\rmfamily\fontsize{9.000000}{10.800000}\selectfont pred: 6}%
\end{pgfscope}%
\begin{pgfscope}%
\definecolor{textcolor}{rgb}{0.000000,0.000000,0.000000}%
\pgfsetstrokecolor{textcolor}%
\pgfsetfillcolor{textcolor}%
\pgftext[x=0.314658in,y=1.158268in,left,base]{\color{textcolor}\rmfamily\fontsize{9.000000}{10.800000}\selectfont BA}%
\end{pgfscope}%
\begin{pgfscope}%
\pgfpathrectangle{\pgfqpoint{0.615180in}{1.009752in}}{\pgfqpoint{0.429317in}{0.429317in}}%
\pgfusepath{clip}%
\pgfsys@transformshift{0.615180in}{1.009752in}%
\pgftext[left,bottom]{\pgfimage[interpolate=true,width=0.430000in,height=0.430000in]{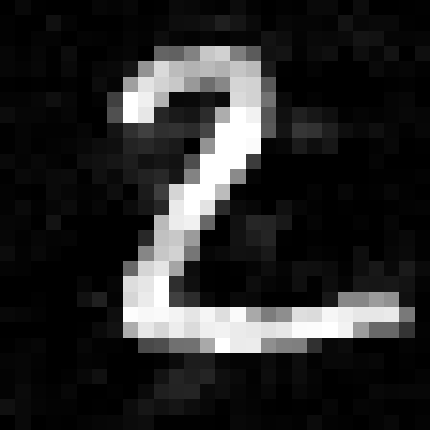}}%
\end{pgfscope}%
\begin{pgfscope}%
\definecolor{textcolor}{rgb}{0.000000,0.000000,0.000000}%
\pgfsetstrokecolor{textcolor}%
\pgfsetfillcolor{textcolor}%
\pgftext[x=0.829839in,y=1.522402in,,base]{\color{textcolor}\rmfamily\fontsize{9.000000}{10.800000}\selectfont pred: 1}%
\end{pgfscope}%
\begin{pgfscope}%
\pgfpathrectangle{\pgfqpoint{1.130360in}{1.009752in}}{\pgfqpoint{0.429317in}{0.429317in}}%
\pgfusepath{clip}%
\pgfsys@transformshift{1.130360in}{1.009752in}%
\pgftext[left,bottom]{\pgfimage[interpolate=true,width=0.430000in,height=0.430000in]{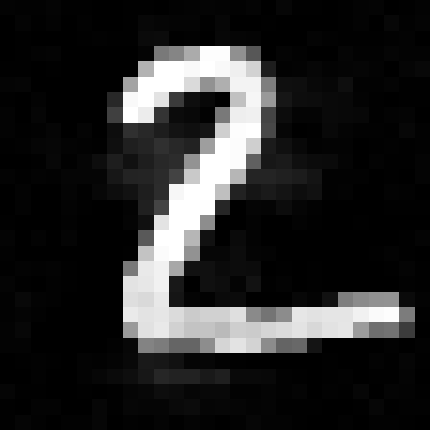}}%
\end{pgfscope}%
\begin{pgfscope}%
\definecolor{textcolor}{rgb}{0.000000,0.000000,0.000000}%
\pgfsetstrokecolor{textcolor}%
\pgfsetfillcolor{textcolor}%
\pgftext[x=1.345019in,y=1.522402in,,base]{\color{textcolor}\rmfamily\fontsize{9.000000}{10.800000}\selectfont pred: 8}%
\end{pgfscope}%
\begin{pgfscope}%
\pgfpathrectangle{\pgfqpoint{1.645540in}{1.009752in}}{\pgfqpoint{0.429317in}{0.429317in}}%
\pgfusepath{clip}%
\pgfsys@transformshift{1.645540in}{1.009752in}%
\pgftext[left,bottom]{\pgfimage[interpolate=true,width=0.430000in,height=0.430000in]{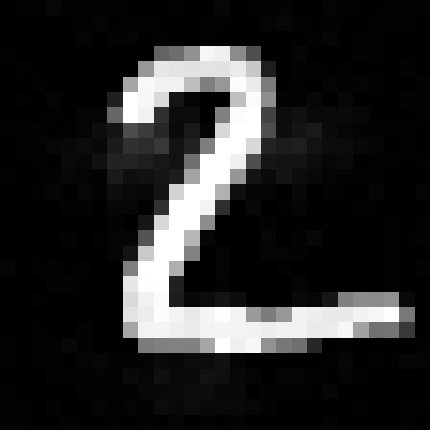}}%
\end{pgfscope}%
\begin{pgfscope}%
\definecolor{textcolor}{rgb}{0.000000,0.000000,0.000000}%
\pgfsetstrokecolor{textcolor}%
\pgfsetfillcolor{textcolor}%
\pgftext[x=1.860199in,y=1.522402in,,base]{\color{textcolor}\rmfamily\fontsize{9.000000}{10.800000}\selectfont pred: 8}%
\end{pgfscope}%
\begin{pgfscope}%
\pgfpathrectangle{\pgfqpoint{2.160721in}{1.009752in}}{\pgfqpoint{0.429317in}{0.429317in}}%
\pgfusepath{clip}%
\pgfsys@transformshift{2.160721in}{1.009752in}%
\pgftext[left,bottom]{\pgfimage[interpolate=true,width=0.430000in,height=0.430000in]{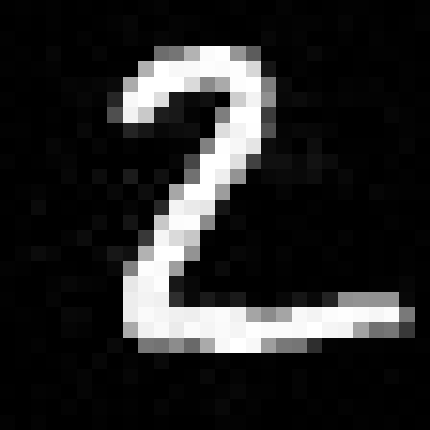}}%
\end{pgfscope}%
\begin{pgfscope}%
\definecolor{textcolor}{rgb}{0.000000,0.000000,0.000000}%
\pgfsetstrokecolor{textcolor}%
\pgfsetfillcolor{textcolor}%
\pgftext[x=2.375379in,y=1.522402in,,base]{\color{textcolor}\rmfamily\fontsize{9.000000}{10.800000}\selectfont pred: 3}%
\end{pgfscope}%
\begin{pgfscope}%
\definecolor{textcolor}{rgb}{0.000000,0.000000,0.000000}%
\pgfsetstrokecolor{textcolor}%
\pgfsetfillcolor{textcolor}%
\pgftext[x=0.314658in,y=0.364567in,left,base]{\color{textcolor}\rmfamily\fontsize{9.000000}{10.800000}\selectfont PGD}%
\end{pgfscope}%
\begin{pgfscope}%
\pgfpathrectangle{\pgfqpoint{0.615180in}{0.216050in}}{\pgfqpoint{0.429317in}{0.429317in}}%
\pgfusepath{clip}%
\pgfsys@transformshift{0.615180in}{0.216050in}%
\pgftext[left,bottom]{\pgfimage[interpolate=true,width=0.430000in,height=0.430000in]{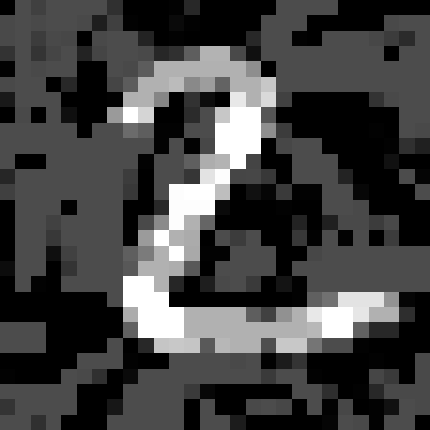}}%
\end{pgfscope}%
\begin{pgfscope}%
\definecolor{textcolor}{rgb}{0.000000,0.000000,0.000000}%
\pgfsetstrokecolor{textcolor}%
\pgfsetfillcolor{textcolor}%
\pgftext[x=0.829839in,y=0.728701in,,base]{\color{textcolor}\rmfamily\fontsize{9.000000}{10.800000}\selectfont pred: 0}%
\end{pgfscope}%
\begin{pgfscope}%
\pgfpathrectangle{\pgfqpoint{1.130360in}{0.216050in}}{\pgfqpoint{0.429317in}{0.429317in}}%
\pgfusepath{clip}%
\pgfsys@transformshift{1.130360in}{0.216050in}%
\pgftext[left,bottom]{\pgfimage[interpolate=true,width=0.430000in,height=0.430000in]{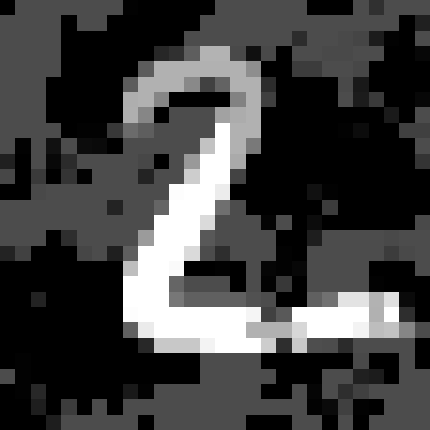}}%
\end{pgfscope}%
\begin{pgfscope}%
\definecolor{textcolor}{rgb}{0.000000,0.000000,0.000000}%
\pgfsetstrokecolor{textcolor}%
\pgfsetfillcolor{textcolor}%
\pgftext[x=1.345019in,y=0.728701in,,base]{\color{textcolor}\rmfamily\fontsize{9.000000}{10.800000}\selectfont pred: 6}%
\end{pgfscope}%
\begin{pgfscope}%
\pgfpathrectangle{\pgfqpoint{1.645540in}{0.216050in}}{\pgfqpoint{0.429317in}{0.429317in}}%
\pgfusepath{clip}%
\pgfsys@transformshift{1.645540in}{0.216050in}%
\pgftext[left,bottom]{\pgfimage[interpolate=true,width=0.430000in,height=0.430000in]{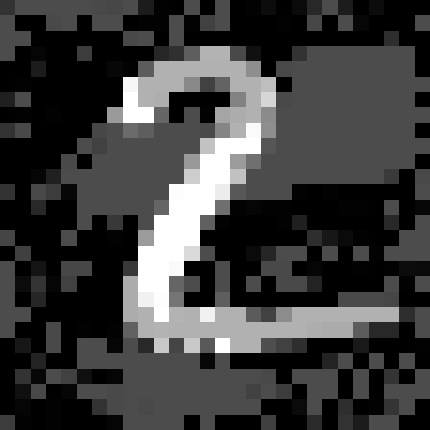}}%
\end{pgfscope}%
\begin{pgfscope}%
\definecolor{textcolor}{rgb}{0.000000,0.000000,0.000000}%
\pgfsetstrokecolor{textcolor}%
\pgfsetfillcolor{textcolor}%
\pgftext[x=1.860199in,y=0.728701in,,base]{\color{textcolor}\rmfamily\fontsize{9.000000}{10.800000}\selectfont pred: 8}%
\end{pgfscope}%
\begin{pgfscope}%
\pgfpathrectangle{\pgfqpoint{2.160721in}{0.216050in}}{\pgfqpoint{0.429317in}{0.429317in}}%
\pgfusepath{clip}%
\pgfsys@transformshift{2.160721in}{0.216050in}%
\pgftext[left,bottom]{\pgfimage[interpolate=true,width=0.430000in,height=0.430000in]{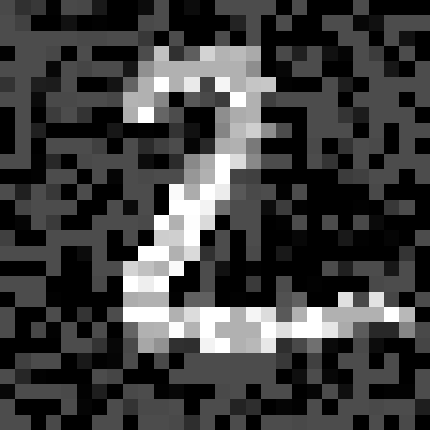}}%
\end{pgfscope}%
\begin{pgfscope}%
\definecolor{textcolor}{rgb}{0.000000,0.000000,0.000000}%
\pgfsetstrokecolor{textcolor}%
\pgfsetfillcolor{textcolor}%
\pgftext[x=2.375379in,y=0.728701in,,base]{\color{textcolor}\rmfamily\fontsize{9.000000}{10.800000}\selectfont pred: 5}%
\end{pgfscope}%
\end{pgfpicture}%
\makeatother%
\endgroup%
    \caption{Adversarial samples generated with CW$_{2}$, BA$_2$, and PGD$_{\infty}$ on the custom architecture models with different privacy settings (from left to right): SGD; $\sigma=1.3, C=1$; $\sigma=3, C=1$; $\sigma=3, C=10$.}
    \label{fig:adv_samples}
\end{figure}

\begin{figure*}
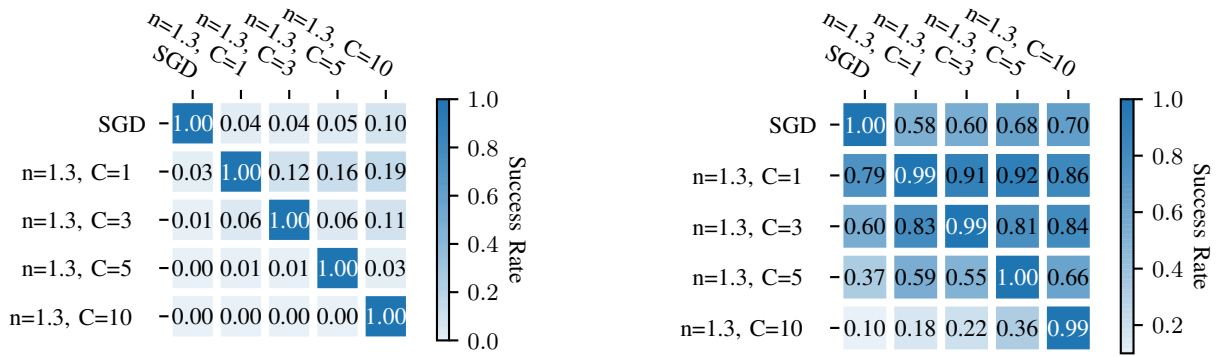

    \centering
    \begin{subfigure}[b]{0.49\textwidth}
         \centering
         \input{imgs/transfer_cw.pgf}
         \caption{CW$_{2}$ attack. For perturbation values per model see Table \ref{tab:cw_epsilon_rates}.}
         \label{fig:adv_samples_trans_1}
     \end{subfigure}
     \begin{subfigure}[b]{0.49\textwidth}
         \centering
         \input{imgs/transfer_linf.pgf}
         \caption{PGD$_{\infty}$ attack. Perturbation $\epsilon=0.3$ and 40 attack iterations.}
         \label{fig:adv_samples_trans_2}
     \end{subfigure}
    \caption{Transferability of generated adversarial examples between custom models. 
    Adversarial examples generated on models with settings depicted in the rows and evaluated against models with settings depicted in the columns.
    High success rates indicate that the adversarial examples transfer well, low success rates indicate low transferability.}
    \label{fig:adv_samples_trans}
\end{figure*}

In addition to evaluating the success of directly attacking the models in a white-box setting, transferability attacks were conducted.
Thereby, the possibility of transferring adversarial examples between DP models with different parameters, or between private and non-private models was quantified (see Figure \ref{fig:adv_samples_trans}).

\paragraph{CW$_{2}$-Transferability}
In the first part of the experiment, the transferability of adversarial examples generated with the CW$_{2}$ attack was evaluated.
For this purpose, for each of the models under attack, first, 1000 correctly classified test samples were chosen randomly.
Then, the CW$_2$ attack was used to craft adversarial examples with 100\% success rate on the respective surrogate models.
Finally, the original model's accuracy on the generated adversarial examples was measured to determine the success rate of the attack.
Figure~\ref{fig:adv_samples_trans_1} depicts the results for the custom model.
Results for the LeNet models look similar and are, therefore, not shown here.

The transferability of adversarial examples created with the CW$_2$ attack might be influenced by the different levels of applied perturbations:
The perturbation budgets that lead to a 100\% success rate in the CW$_2$ attack vary between models and tend to be lower for DP models (see Table~\ref{tab:cw_epsilon_rates}).

\paragraph{PGD$_{\infty}$-Transferability}
Therefore, in the second experiment, the transferability of adversarial examples generated with PGD$_{\infty}$ and a constant perturbation of $\epsilon=0.3$ was assessed.
The procedure of determining the success rates was the same as for the CW$_2$ attack.
Figure~\ref{fig:adv_samples_trans_2} summarizes the results of this test on the custom models.
Again, results on LeNet models were similar and are, therefore, not shown here.

The experiments suggest that for PGD$_{\infty}$, the adversarial examples transfer significantly better than for CW$_2$.
Still, the same trends can be observed for both scenarios:
Adversarial examples seem to transfer less well from DP to non-DP models than the other way round.
Additionally, adversarial examples generated on models with higher clip norms seem to transfer less well to other models than adversarial examples generated on models with lower clip norms.
The adversarial examples generated on DP models with smaller clip norms ($C=1$ and $C=3$) transfer better to other DP models than the adversarial examples generated on the non-private baseline models.
Also, models trained with higher clip norms exhibit an increased vulnerability to transferability attacks in comparison to models with lower clip norms.

These results suggest that transfer attacks between models with different privacy settings are indeed successful.
Again, the models that caused plateaus in the success rate when directly executing PGD$_{\infty}$ against them seem to be the most vulnerable ones, according to this experiment.
This is another indicator for their apparent robustness being just a consequence of low-utility gradients. 
Another interesting observation is the fact that adversarial examples tend to transfer better between DP models than from non-DP to DP models.

To investigate this effect further, adversarial examples created with the three considered attack methods on models with different privacy settings were visually examined (see Figure~\ref{fig:adv_samples}). 
The first column of the figures shows adversarial examples generated for the non-private baseline model.
The results of the previous section and a visual inspection of the generated samples by CW$_2$ and BA$_2$ suggest that the more private the models are, the less perturbation budget is required to fool the model with a 100\% success rates.

The adversarial examples displayed for the PGD$_{\infty}$ method all have the same perturbation budget of $\epsilon=0.3$.  
Interestingly, the generated samples look significantly different between DP and non-DP models. 
Whereas in the non-DP model, the artifacts introduced into the images are grouped into regions, the higher the privacy gets, the more they resemble random noise and are not grouped into regions anymore. 
The finding of this visual inspection also reflect in the evaluation of the transferability experiments.

\section{DP and Gradient Masking}
\label{sec:obfuscation}

\begin{figure}
    \centering
     \input{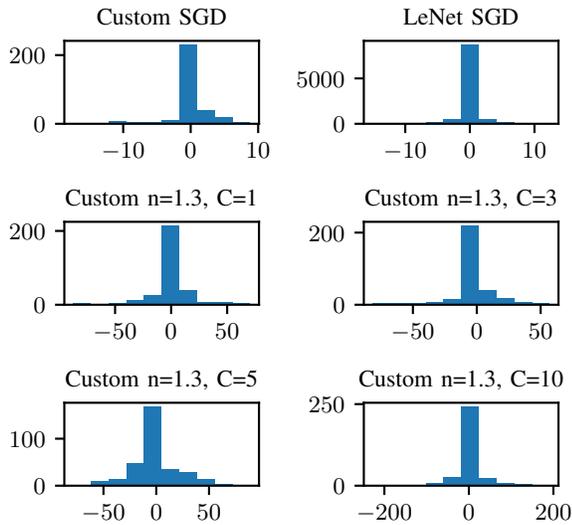}

    \caption{Distribution of the gradient magnitudes at the models' first dense layer. Upper row: baseline models without privacy, lower rows: custom models with DP.}
    \label{fig:gradients}
\end{figure}

\begin{figure}
    \centering
    \begin{subfigure}[b]{0.45\textwidth}
         \centering
         \input{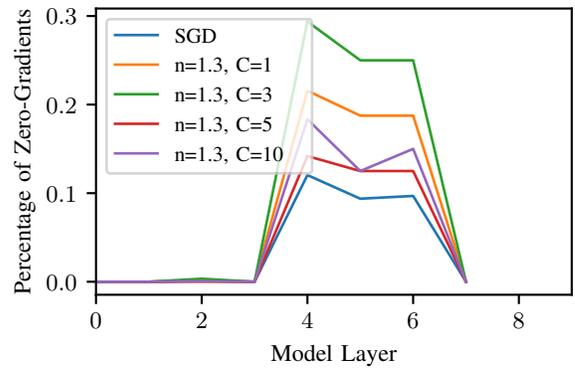}
         \caption{Custom architecture.}
         \label{fig:zero_gradients_1}
     \end{subfigure}
     \begin{subfigure}[b]{0.45\textwidth}
         \centering
         \input{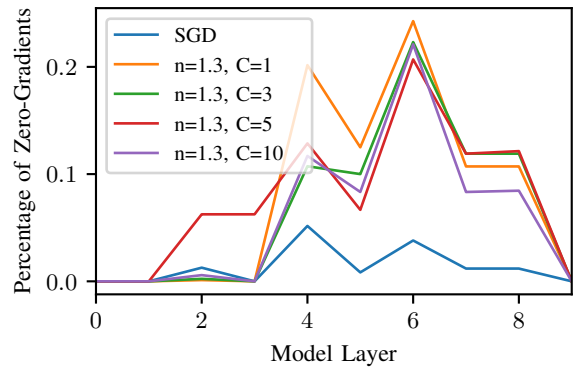}
         \caption{LeNet architecture.}
         \label{fig:zero_gradients_2}
     \end{subfigure}
      
    \caption{Percentage of zero gradients over all model layers.}
    \label{fig:zero_gradients}
\end{figure}

The results presented in \Cref{sec:robustness} and \Cref{sec:transferability} suggest, that DP models may not be generally more robust compared to non-private ones.
Even though the PGD$_{\infty}$ attack did not yield a 100\% success rate, adversaries using the CW$_{2}$ and BA$_2$ were able to fool the evaluated models.
The fact that the gradient-based PGD$_{\infty}$ attack did not achieve a 100\% success rate for certain settings, whereas the other attacks did, suggests instabilities of the models' gradients.
Therefore, two properties of the gradients, namely, their magnitudes and the percentage of zero-gradients were investigated.

Figure~\ref{fig:gradients} displays the gradient magnitudes for both model architectures after the first dense layer for a test data batch.
While the magnitudes of the non-DP models' gradients range from -10 to 10, the range for custom architecture DP models spans from -200 to 200.
With an increasing clip value, the magnitude of the gradients increases.
The same trend, even though not so prominent, can be observed for the LeNet models.

The percentage of zero-gradients per model layer is depicted in Figure~\ref{fig:zero_gradients}.
For DP models, the proportion of gradients with zero magnitude is higher than for non-private models.
For the custom architecture, an increase of this proportion with increasing clip norms can be observed.
In the LeNet models, there are no such significant differences between different privacy parameters.

The observed properties of the gradients in the DP models might explain why PGD$_{\infty}$ did not achieve a 100\% success rate on the models with high noise and high clip norms. 
Due to the masked gradients and potential numerical instabilities during the calculations, no useful information required for a successful attack were available.

\begin{table}[t]
    \centering
    \begin{tabular}{lcc}
    \toprule
    Model & Parameters & Adv. Perturbation \\
    \midrule 
    bs1 (Custom) &	SGD $, epochs=50$& $\epsilon=1.20$\\ 
    bs2 (LeNet) & SGD $, epochs=50$ & $\epsilon=1.21$\\ 
    m1 (Custom)& $\sigma=2, C=5, epochs=50$ & $\epsilon=1.34$\\ 
    m2 (Custom)& $\sigma=2, C=6, epochs=40$ & $\epsilon=1.64$ 	\\ 
    m3 (Custom)& $\sigma=2, C=7, epochs=20$ & $\epsilon=1.82$ 	\\
    \bottomrule
    \end{tabular}
    \caption{Models with masked gradient and the adversarial perturbation that is required to achieve a 100\% success rate with the CW$_2$ attack and \num{20000} attack iterations.}
    \label{tab:settings_obfuscated_models}
\end{table}

\begin{table}[t]
    \centering
    \begin{tabular}{lccc}
    \toprule
    Model & Parameters & \multicolumn{2}{c}{Success Rate} \\
    & &$\epsilon=1$ & $\epsilon=2$\\
    \midrule 
    bs1 (Custom) & SGD & 19.1\% &83.2\%\\ 
    bs2 (LeNet)  &SGD & 27.5\%&79.4\%\\ 
    m1 (Custom) & $\sigma=2, C=5, epochs=50$&95.3\%&99.7\%\\ 
    m2 (Custom)&  $\sigma=2, C=6, epochs=40$ &80.7\%&	95.2\%\\ 
    m3 (Custom)& $\sigma=2, C=7, epochs=20$ &82.7\%&99.8\%	\\
    \bottomrule
    \end{tabular}
    \caption{Success rates of BA$_2$ with \num{25000} attack iterations and different perturbation values on models with masked gradients.}
    \label{tab:boundary_obfuscated_models}
\end{table}

Further investigations even suggest that it is possible to intentionally choose the DP settings in a way that the models still achieve a relatively high accuracy, but cause gradient masking according to the criteria presented in Section~\ref{ssec:rel_grad}.
In general, gradient masking in DP seems to be due to an unfavorable combination of noise, high clip norms, and the model architecture. 
The higher the clip norm, less noise or training epochs already lead to masked gradients in certain architectures. 
In the following, three such exemplary models are depicted and analyzed according to the aforementioned criteria. 
See Table~\ref{tab:settings_obfuscated_models} for their privacy parameter settings and training epochs.

\begin{figure*}
    \centering
    \begin{subfigure}[b]{0.45\textwidth}
         \centering
         \input{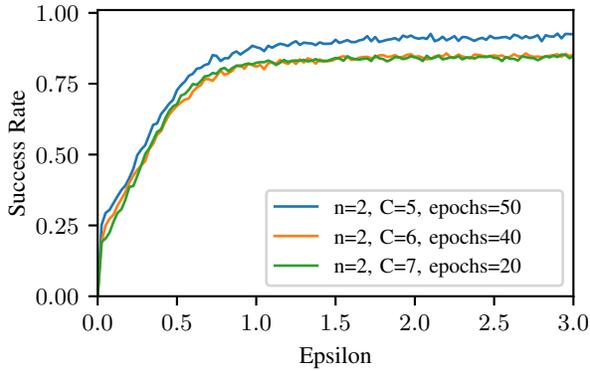}
         \caption{PGD$_{\infty}$: adversarial success rate plotted against adversarial perturbation $\epsilon$.}
         \label{fig:dpobf_exp_1}
     \end{subfigure}
     \begin{subfigure}[b]{0.45\textwidth}
         \centering
         \input{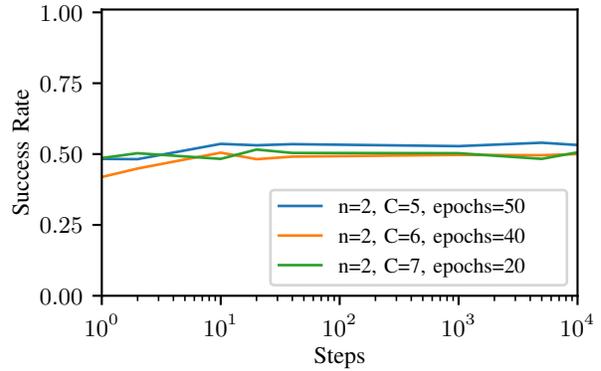}
         \caption{PGD$_{\infty}$: adversarial success rate plotted against number of iterations.}
         \label{fig:dpobf_exp_2}
    \end{subfigure}
    \caption{PGD$_{\infty}$ and BA$_2$ applied to the three DP models depicted in Table~\ref{tab:settings_obfuscated_models}.}
    \label{fig:dpobf_exp}
\end{figure*}


For the models, increasing the number of attack steps, or increasing the adversarial perturbation does not yield an increase in success rate after a certain plateau is reached. 
See Figure~\ref{fig:dpobf_exp} for the experimental results. 
This can be an indication for masked gradients following criterion (5) and (6).

Furthermore, the black-box and gradient-free BA$_2$ achieves similar and higher success rates than the white-box and gradient-based PGD$_{\infty}$ method (see Table~\ref{tab:boundary_obfuscated_models}). 
According to criterion (2) and (3), this might be another indicator for masked gradients.

\begin{figure}
    \centering
         \input{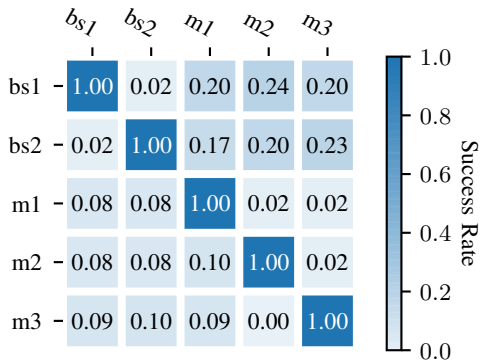}
    \caption{CW$_{2}$ transferability between baseline and models with masked gradients.}
    \label{fig:dpobf_trans}
\end{figure}

Finally, transferability attacks using CW$_{2}$ with 20,000 steps and adversarial perturbation values depicted in Table~\ref{tab:settings_obfuscated_models} were conducted.
See Figure~\ref{fig:dpobf_trans} for the results.
The transferability among the models is significantly higher than the transferability of the models studied in  Section~\ref{sec:transferability} (see Figure~\ref{fig:adv_samples_trans_1}).
Therefore, following criterion (7), the high robustness implied by the success of the PGD$_{\infty}$ attack (see Figure~\ref{fig:dpobf_exp_1} and \ref{fig:dpobf_exp_2}) might be due to gradient masking.

To investigate whether unbounded attacks reach a 100\% adversarial success rate, the CW$_{2}$ attack was conducted.
After 20,000 iterations and with perturbation budgets depicted in Table~\ref{tab:settings_obfuscated_models}, the attack achieved a 100\% success rate for each of the three models.
Therefore, criterion (4) is met.
Interestingly, several of the generated samples for model m3 are entirely black.
Carlini \etal~\cite{Carlini.2019On} suggest that this kind of behavior might occur when the only possibility to successfully fool a model is to actually turn an instance  into one of the other class.

\begin{figure}
    \centering
         \input{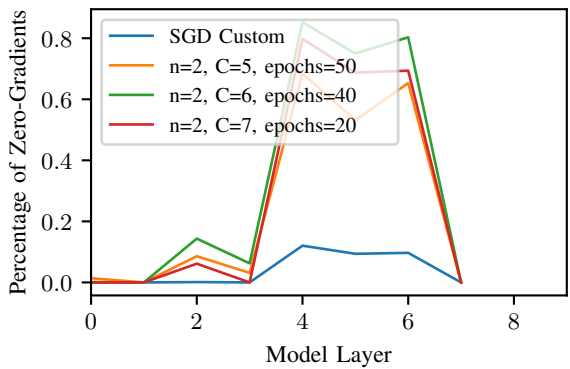}
    \caption{Zero gradients over all model layers.}
    \label{fig:dpobf_zero_gradients_perc}
\end{figure}

\begin{figure*}
    \centering
         \input{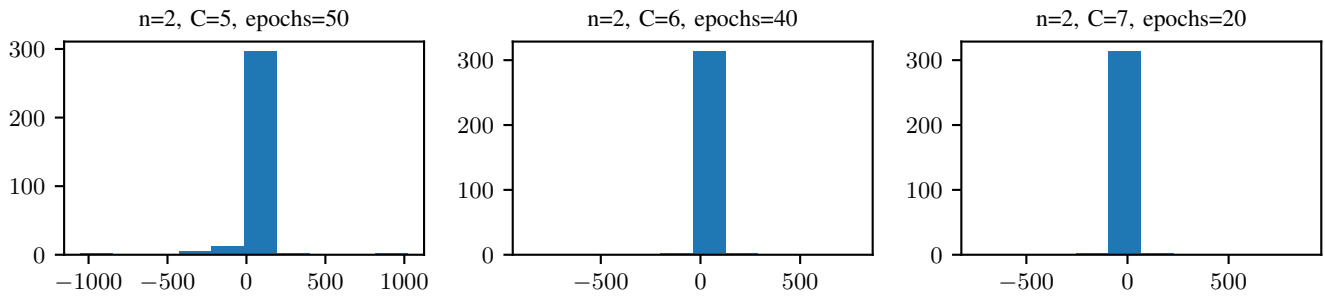}
    \caption{Gradient magnitudes at model's first dense layer.}
    \label{fig:dpobf_zero_gradients}
\end{figure*}

Looking into the gradients of the three models reveals that the majority of the them is zero (see Figure~\ref{fig:dpobf_zero_gradients_perc}), but that the magnitudes of the remaining gradients are extremely high (see Figure \ref{fig:dpobf_zero_gradients}).
This is another indicator for reduced gradient usefulness due to gradient masking and might explain, why crafting adversarial examples based on the model gradients yields little success.

\section{Discussion and Outlook}
\label{sec:discussion}

The experiments in this paper have shown that DP models exhibit an increased adversarial vulnerability in comparison to non-DP models.
Furthermore, the observations by Tursynbek \etal~\cite{Tursynbek.2020Robustness} (some DP parameter combinations with high clip norms yield high robustness) are likely due to gradient masking. 

There are several  possible explanations for the differences in robustness between DP and the baseline models.
According to Demontis \etal~\cite{Demontis.2019Why} the larger the gradients in a target model, the larger the impact of an adversarial attack.
As the experiments have shown, DP models' gradients are much larger than normal models' gradients, potentially explaining their increased vulnerability. 
To counteract this factor, it might be helpful to regularize the gradients in DP model training.

Another reason for the increased vulnerability of DP models might be their decision boundaries.
Tursynbek \etal~\cite{Tursynbek.2020Robustness} show that training with DP-SGD affects the underlying geometry of the decision boundaries.
In their example, it becomes visible that the DP-SGD training results in more fragmented and smaller decision regions. This increases the chances to generate an adversarial example with less perturbation.

In a similar vain, Demontis \etal~\cite{Demontis.2019Why} suggest that the loss surface of a model has an influence on the robustness against adversarial examples.
They state that if the landscape of a model is very variable, it is much likely that slight changes to data points will encourage a change in the local optima of the corresponding optimization problem.
As a consequence the authors conclude that attack points might not transfer correctly to another model with a potentially more stable landscape.
The experiments of this work depict that adversarial examples generated on DP models transfer less to normal models than the other way round, and that adversarial examples crafted on models with higher clip norms transfer less than the ones from models with lower clip norms.
This might also be due to the models' loss landscapes.
Future work could, therefore, investigate the loss surface of the DP models more thoroughly.

Previous results by Papernot \etal~\cite{Papernot.2020Tempered} suggest that applying DP in combination with standard RELU functions might lead to exploding activations in the resulting models.
The authors suggest using sigmoid activation functions to counteract this effect and to, thereby, improve the training process and achieve higher accuracy scores.
In future work, it would be interesting to investigate whether this replacement of the activation functions might also be beneficial for the model robustness.
The authors also conclude that the models' activation functions have the largest influence on the success of DP training.
However, the experiments in this work suggest, that the LeNet architecture might be more robust, with less zero-gradients and lower magnitudes of the remaining gradients.
Hence, when considering privacy in combination with security, the model architecture might be an important factor to consider as well \cite{Su.2018Is}.



The results of this work raise the question whether training DNNs with DP does necessarily cause an increase in model vulnerability against adversarial examples.
For the current state, the experiments suggest that achieving privacy does have a negative impact on model robustness.
At the same time, this work also highlights a direction of research that might be worth pursuing in the future, namely controlling the gradients in DP model training.
Pinot \etal~\cite{Pinot.2019unified} show that, in principle, Renyi-DP~\cite{Mironov.2017Renyi}, which is used in the DP-SGD, and adversarial robustness share equivalent goals.
Therefore, future work could investigate how DP training can be adapted to simultaneously improve robustness and which factors, apart from the gradients, cause current DP model's vulnerability.

\section{Conclusion}
\label{sec:conclusion}
Making DNNs more private and more robust are important tasks that have long been considered separately.
However, to solve both problems at the same time, it is beneficial to understand which impact they have on each other.
This work addressed this question from a privacy perspective, evaluating how training DNNs with DP-SGD affects the robustness of the models.
The experiments demonstrated that DP-SGD training causes a decrease in model robustness.
By conducting a broad range of adversarial attacks against DP models, it was shown that the positive effects of DP-SGD observed in previous work might be largely due to gradient masking, and therefore, provide a wrong sense of security.
Analyzing and comparing the gradients of DP and non-DP models demonstrated that DP training leads to larger gradient magnitudes and more zero gradients, which might be a reason for the DP models' higher vulnerability.
Finally, this work is the first to depict that certain DP parameter combinations in the DP-SGD can intentionally cause gradient masking.
As a consequence, future work may further investigate the influence of DP training on the models.
This could serve as a basis during parameter and architecture selection such that private training does not oppose the goal of security and can, hence, be applied also in critical scenarios.

\printbibliography

\end{document}